\documentclass[multphys,vecphys]{svmult}

\usepackage{makeidx}     
\usepackage{graphicx}    
\usepackage{multicol}    

\makeindex             

\begin{document}

\title*{Galactic Center Molecular Clouds}
\author{Rolf G\"usten\inst{1} \and
Sabine D. Philipp\inst{2}}
\institute{MPI f\"ur Radioastronomie, Auf dem H\"ugel 69, 53121 Bonn, Germany
\texttt{rguesten@mpifr-bonn.mpg.de}
\and
MPI f\"ur Radioastronomie, Auf dem H\"ugel 69, 53121 Bonn, Germany
\texttt{sphilipp@mpifr-bonn.mpg.de}}
\maketitle
\section{Introduction}
\label{sec:1}
At a distance of 8 kpc only, the center of the Galaxy provides a unique testbed for studies about the physics of the interstellar medium and star formation in the
nuclei of galaxies. Since it shares many properties of other, more spectacular nuclei, a detailed investigation of the Galactic Center is a necessary
step to gain better understanding of the physical processes governing galactic nuclei in general. \\
We review the  characteristics of the central molecular cloud layer, the physical properties of these clouds, and their interplay with
the magnetic field.
\section{The Galactic Center Molecular Clouds}
\label{sec:2}
\subsection{Large-Scale Distribution: Overview}
\label{sec:1}
The Galactic Center is characterized by a large concentration of dominantly molecular gas ($\sim$10$^{8}M_{\odot}$), associated with the innermost
few hundred parsec of the Galaxy (the size of the molecular layer is 450 x 50 pc). Despite its exceptionally high gas surface densities ({\small{$\geq$}}100 M$_{\odot}$pc$^{-2}$),
the Galactic Center is highly depleted, with only a small fraction of matter left in the interstellar gas phase, $\mu \sim$ 0.01. The high degree of astration
is reflected in the enhanced chemical enrichment of the gas with ejecta from evolved stars (Table 2).
\\
The spatial distribution of the molecular layer about the nucleus is
asymmetric ($\sim$2/3 displaced towards positive longitudes), with the gas column densities peaking towards the Sgr B2 complex (see \cite{bal85} for an early data base).
Much of the mass is concentrated in cloud complexes as large as 50-70 pc.   \\
Already back in the 80ies, molecular excitation studies have revealed the exceptional physical properties of these clouds: compared with clouds
in the Galactic disk, GC clouds are denser, warmer, more turbulent and more massive. More recent surveys of the warm CO and also the atomic fine structure lines extended well into the
submm-regime (\cite{ojh01}\cite{kim02}\cite{mar03} with AST/RO, \cite{phi04} with CHAMP at the CSO; see Table 1), thus better constraining the thermodynamics of the gas. Similarly, technology has made possible to develop
large submm bolometer arrays, which allowed mapping the large-scale distribution of the warm dust (and more recently, its polarisation) with $\sim$10$"$ resolution across the central layer
(\cite{zyl98}\cite{pie00}).
\small{
\begin{table}[t]\centering
\begin{tabular}{lcccl}
\hline \hline \\[-2ex]
 \bf{Tracer}      &\multispan2 \hspace*{1mm}\bf{Surveyed area} $[$deg$]$ \hfil & \bf{Beam} & \bf{Reference} \\
       &\bf{Longitude} & \bf{Latitude}         &   &       \\[0.5ex]
\hline    \\[-2ex]
$^{13}$CO(1-0)\hspace*{5mm}     & -0.8 to 1.7   & -0.35 to 0.35 & \hspace*{8mm}17$"$\hspace*{8mm} &Oka et al. 1998 \\
$^{12}$CO(1-0)     & -1.5 to 3.47  & -0.6 to 0.6   & 16$"$ &Oka et al. 1998 \\
$^{13}$CO(2-1)     & -0.23 to 0.27 & -0.17 to 0.20 & 12$"$  &Philipp et al. 2004 \\
$^{12}$CO(2-1)     & -6.0 to 6.0   & -2.0 to 2.0   & 540$"$ &Sawada et al. 2001 \\
$^{12}$CO(4-3)     & -0.23 to 0.32 & -0.17 to 0.20 & 16$"$  &Philipp et al. 2004 \\
                  & -1.3 to 2.0   & -0.3 to 0.2   & 103$"$ &Martin et al. 2003 \\
$^{12}$CO(7-6)     & -1.3 to 2.0   & -0.3 to 0.2   & 58$"$  &Martin et al. 2003 \\[0.5ex]
\hline  \\[-2ex]
CS(1-0)     & -1.1 to 1.7  & -0.3 to 0.2   & 34$"$ &Tsuboi et al. 1999 \\
CS(2-1)     & -0.23 to 0.20 & -0.17 to 0.20 & 25$"$  &Philipp et al. 2004 \\
CS(3-2)     & -0.23 to 0.20 & -0.17 to 0.20 & 17$"$  &Philipp et al. 2004 \\
\hline  \\[-2ex]
CI($^{3}$P$_1$-$^{3}$P$_0$) &  -0.10 to 0.27    & -0.08 to 0.17     & 15$"$ & Philipp et al. 2004  \\
			    &      &             & 103$"$ & Martin et al. 2003  \\[0.5ex]
\hline \hline  \\[-1ex]
C  1200 $\mu$m$$             & -0.5 to 0.8 & -0.2 to 0.2  & 11$"$ & Zylka et al. 1998 \\
C  850 $\mu$m$$       & -1.0 to 1.7 & -0.3 to 0.2 &  15$"$ & Pierce-Price et al. 2000 \\
C  450 $\mu$m$$      &   -1.0 to 1.7 & -0.3 to 0.2 &  8$"$ & Pierce-Price et al. 2000 \\[0.5ex]
 \hline \hline  \\[-1ex]
\end{tabular}
\caption{Compilation of some of the more recent mm/submm line and continuum surveys towards the Galactic Center. The area surveyed and the angular resolution of the observations is given.}
\end{table}
}
\begin{figure}[h]
\setlength{\unitlength}{1mm}
\centering
\begin{picture}(200,145)
\includegraphics{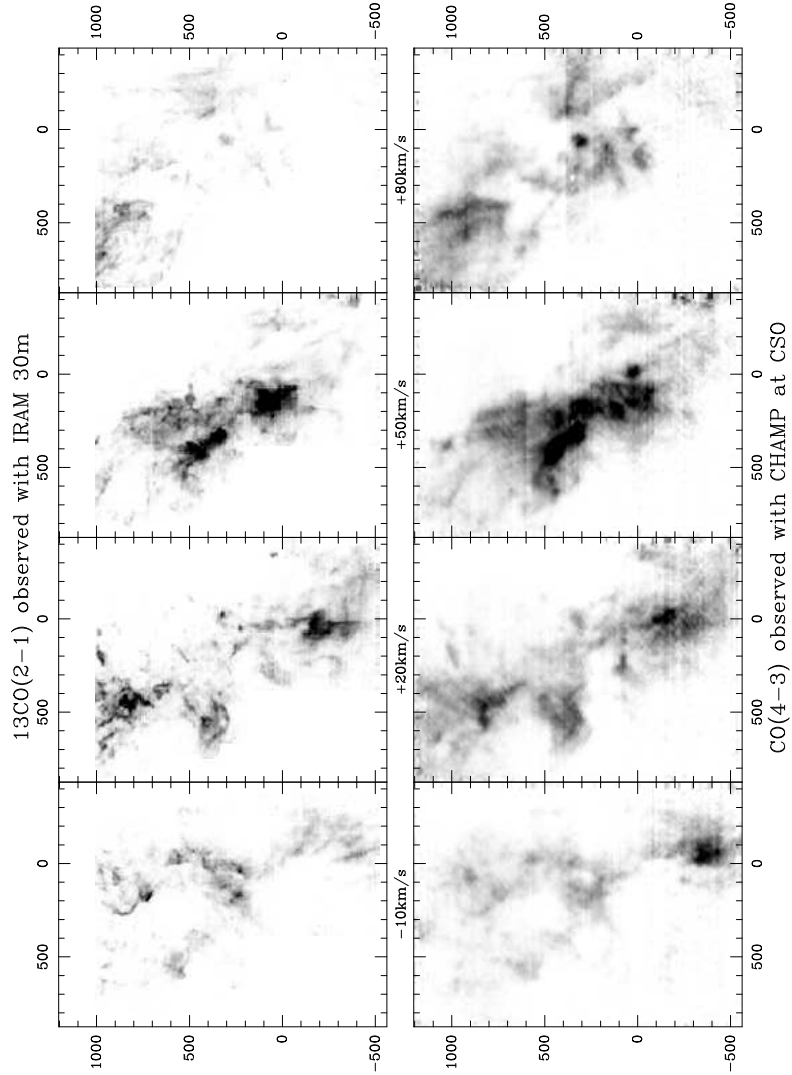}
\end{picture}
\caption{Selected channels maps of $^{13}$CO(J=2-1) and CO(J=4-3) velocity-integrated brightness temperatures, revealing the tremendous
morphological fine-structure visible in high-resolution observations \cite{phi04}. Offsets in arcsec rel. to Sgr A$^*$. The data have
 been sampled into 5 km/s wide velocity bins. The observations were performed with the IRAM 30m telescope (11")
and with CHAMP at the CSO (15" resolution).}
\label{fig_sgra_co}
\end{figure}
\begin{figure}[hbt]
\setlength{\unitlength}{1mm}
\begin{picture}(200,35)
\end{picture}
\caption{Submm dust emission at 850 $\mu$m as measured with the JCMT/SCUBA array with 15" angular resolution \cite{pie00}.
Prominent GC cloud complexes are labelled. \emph{Version of this article including this figure can be obtained from
\texttt{ftp://ftp.mpifr-bonn.mpg.de/outgoing/sphilipp}.}}
\label{fig_scuba}
\end{figure}
\subsection{Physical Characteristics of the GC Cloud Population}
\label{sec:2}
The average characteristics of the molecular clouds in the Galactic Center differ significantly from those in the outer Galaxy: elevated gas
temperatures ($<$T$_{kin}$$> \sim$50-70 K), high densities ($<$n$>$ $\sim$10$^4$ cm$^{-3}$) and strong turbulence, usually found in compact star forming dense cloud
cores only, are pervasive and likely a consequence of tidal friction in the steep gravitational potential of the central bulge: clouds must be
more massive in order to survive against the strong tidal forces. Dissipation of the high supersonic internal velocities (V $\sim$15-30 km s$^{-1}$),
comparable to the intercloud velocity dispersion, is a likely source for the heating of the molecular gas.
Evidence for a pervasive magnetic field component with milliGauss field strength throughout the Galactic Center region is discussed in Sect.3.
\subsubsection{Statistical Properties: Velocity Dispersion and Mass Spectrum}
Based of the recent high-resolution line surveys of CS(1-0) \cite{tsu99} and CO(1-0) \cite{oka98a} (Table 1), a structural decomposition \cite{stu90} and statistical
analysis of cloud properties was performed \cite{oka01}\cite{miy00}.
The velocity linewidth - size dependence of GC clouds basically follow the same relation as disk clouds, $\bigtriangleup$V $\propto$R${^\alpha}$,
with ${\alpha}$ $\sim$0.4-0.5, but they are about five times more turbulent (Fig. 3, left panel). The mass spectrum (M $\geq$10$^{4}$ $M_{\odot}$)
is well represented by a power law, dN/dM $\propto$M$^{-\gamma}$ with $\gamma$=1.6$\pm$0.1, very similar again to indices derived for Galactic disk cloud complexes.
Primarily due to the large velocity dispersion, virial masses exceed {\it{observed}} masses (derived from the line intensities, assuming LTE) by an
order of magnitude (Fig.3, right panel). The discussion about what stabilizes these clouds is ongoing and controversial. External pressure from a hot plasma is
limited observationally to P$_{pla}$ $\sim$10$^{-9.2}$ erg/cm$^{3}$, and falls short by an order of magnitude to the required pressure
(P$_{turb}$ $\sim$10$^{-8}$ erg/cm$^{3}$). A strong magnetic field could bind the cloud: assuming equilibrium between turbulent and magnetic pressure, B$^2$/8$\pi$,
stabilizing intra-cloud fields of $\sim$0.5 mG are required (such field strengths are observationally not excluded, see Sect.3).
\begin{figure}[hb]\centering
\setlength{\unitlength}{1mm}
\begin{picture}(200,55)
\includegraphics{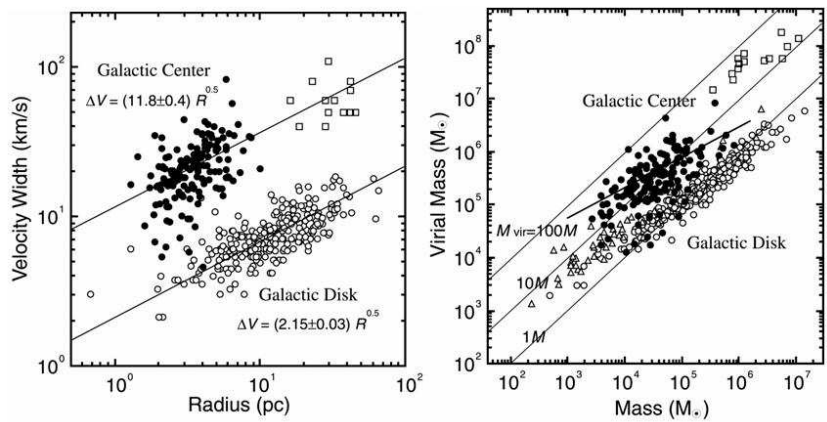}
\end{picture}
\caption{Properties of the Galactic Center clouds \cite{miy00}, derived from structural decomposition of the CS(1-0)\cite{tsu99} and CO(1-0)\cite{oka98a} survey data.
The intracloud velocity dispersion and the virial mass of the clumps are displayed, in comparison with findings for normal Galactic disk clouds (open circles).}
\label{fig_clouds}
\end{figure}
\\[-8ex]
\subsubsection{Temperatures and Densities}
A number of studies with density-sensitive molecular species (like CS \cite{bal85}, H$_2$CO \cite{gue83}, HC$_3$N \cite{wal86}) have shown that average gas densities
of order 10$^4$ cm$^{-3}$ are characteristic for GC clouds. Detailed excitation studies suggest a clumpy morphology with higher density clumps ($\sim$10$^5$ cm$^{-3}$)
embedded in a lower density interclump medium ($\sim$10$^{3.7}\ $cm$^{-3}$) \cite{wal86}. High densities of order 10$^{4}$$\cdot$(75 pc/R$_{gc}$)$^{1.8}$  [cm$^{-3}$]
 have been argued necessary for
a cloud at galactocentric distance R$_{gc}$ to survive the tidal stresses in the steep Galactic Center potential.  \\
 High kinetic temperatures of the gas have first been deduced from the metastable transitions of ammonia NH$_3$ \cite{gue81}\cite{mor83}\cite{hue93}. Warm gas
  with elaborated bulk temperatures ($\sim$50 K) has
been detected towards all the clouds surveyed. The new large-scale AST/RO maps in the submm CO lines confirm these
findings \cite{kim02}\cite{mar03}. Soon however, it was noticed that towards all line-of-sights there is evidence
 in the excitation studies for co-existing gas phases covering a much wider range of temperatures. This is best reflected in the H$_2$ S(0) to S(5) pure-rotational
 line studies performed with the SWS on board of
 ISO \cite{rod01}, covering excitation energies above ground up to a few 1000 K (see their Fig.5). Their dedicated excitation analysis re-assures the earlier findings:
 warm temperatures are pervasive in galactic center clouds and not necessarily related to star formation activity, and there is a wide hierarchy of temperatures:
 while temperatures traced by the lower S-transitions measure T$_{32}$ $\sim$150 K, the high-J transitions yield T$_{76}$ $\sim$600 K but derive from only $\sim$1$\%$
 of the gas column density residing at 150 K.
 \\
Such pervasive high gas temperatures have stimulated from the very beginning a lively discussion about the underlying heating mechanisms. Direct heating by collisional
dissipation of energy via dust particles has soon been eliminated because of the uniformly low temperature of the dust in the Galactic Center (T$_{d}$ = 21$\pm$2 K $\ll$ T$_{gas}$ \cite{pie00}).
So the quest is up for a heating mechanisms that acts directly on the gas particles. A number of scenarios have been proposed that potentially could
deliver the energy required to compensate for the cooling rate of the gas
(the combined H$_2$ and CO cooling rate is estimated to $\rm\Lambda \sim$6$\times$10$^{-22}$ erg s$^{-1}$cm$^{-3}$ \cite{rod01}): magnetic viscous heating, though requiring mG fields \cite{gue81},
small scale dissipation of supersonic turbulence \cite{wil82}\cite{gue85}, large(r) scale J- and C-shocks as well as UV-heating in exposed photo-dissociation layers.
Most likely, all of these processes are on stage:  \\[-3.5ex]
\begin{list}{$\bullet$}{\leftmargin0.5cm}
\item
towards M-0.96+0.13, \cite{rod01} find evidence for large-scale shock heating in their H$_2$-data,
\item
the molecular gas associated with the Arched Filaments (north of SgrA) and with G0.18-0.04 \underline{is} known to be exposed to a strong UV field from the Quintuplet
\cite{phi04b} and Arches clusters, respectively (see below),
\item
for the majority of clouds, however, dissipation of supersonic turbulence seems the most attractive heating process. Following Black \cite{bla87} the
heating rate can be estimated as
\begin{equation}
\rm\Gamma \sim 3.5 \cdot 10^{28 }\,\, v_t^3\cdot n(\rm{H_2})\cdot(1\  pc/\ell) \hspace*{15mm}  [erg \ s^{-1}cm^{-3}].
\end{equation}
With $\ell \sim$5 pc, v$_t$ $\sim$15 km\ s$^{-1}$, n(H$_2$) $\sim$10$^3$cm$^{-3}$ (as
typical for GC clouds), we calculate $\rm\Gamma$ $\sim$5$\cdot$10$^{-22}$ erg s$^{-1}$cm$^{-3}$, comparable to the cooling rate of the gas.
\end{list}
\subsubsection{Star Formation in the Galactic Center}
In view of these dramatically different initial conditions, star formation proceeds at a surprisingly normal rate (better said, the formation of
massive stars, because we have limited insight into the IMF). For a normal IMF, the overall star formation rate across
the central layer is $\rm\Phi$ $\sim$0.3-0.6 M$_{\odot}$/yr \cite{gue89}, about 10$\%$ of the galactic star formation rate, with an efficiency
$\rm\Phi /M_g$  $\sim$5$\times10^{-9}$yr$^{-1}$ that compares well with the average galactic rate (Table 2). Details are beyond the scope of this review
(see e.g. \cite{mor96}\cite{fig01}), but there is strong evidence that the peculiar boundary conditions of the GC clouds do affect the dominant
mode of star formation: the tidal forces, the large turbulence and, likely, the strong magnetic fields are expected to favor the formation of massive
stars. And in fact, the most spectacular young stellar clusters in the Galaxy (the Quintuplet cluster illuminating G0.18-0.04 \cite{fig99}, and
the Arches cluster near the thermal Arched Filaments \cite{fig02}, the Central cluster) are all found
in the central few 10 parsecs. The Arches cluster compares in age ($2.5\cdot 10^{6}$yr) and content ($>$150 O-stars, 4$\cdot10^{51\ }$N$_{lyc}$ s$^{-1}$)
only to R136 in 30 Dor.
\begin{table}[ht]
\begin{tabular}{lcc}
\hline
\\
                                   & \bf{\large {Central 500 pc}} & \bf{\large {Galactic Disk}} \\[1ex]
\bf{Masses and Densities} &  & \\
\\[-2ex]
\vspace*{3mm}
Stars $[M_{\star}]$                 & 10$^{9.8}M_{\odot}$         & -- --  \\
Gas, atomic                         & 10$^{6.4}M_{\odot}$         & $\sim 10^{9}M_{\odot}$  \\
\vspace*{3mm}
\dots \hspace*{2mm} molecular       & 10$^{7.9}M_{\odot}$         & $\sim 10^{9}M_{\odot}$  \\
Gas fraction $\mu=M_g/M_{\star}$\hspace*{5mm}    & $\sim 0.01$                 & $\sim 0.05 - 0.10$ \\
Fract. abundance $[$HI$]/[$H$_2]$          & $\sim 0.05$                 & $\sim 2$ \\
Gas density $<$n$>_{\rm{v}}$               & 100 cm$^{-3}$               & 1 -- 2  cm$^{-3}$ \\
\vspace*{3mm}
Gas surface density $\sigma$    & $\ge 100M_{\odot}$pc$^{-2}$ & $\sim 5M_{\odot}$pc$^{-2}$ \\
\\[-2ex]
\bf{Star Formation Rate} &  & \\
\\[-2ex]
Rate ${\rm{\Phi}}$                         & 0.3 -- 0.6$M_{\odot}$yr$^{-1}$ & $\sim 5.5 M_{\odot}$yr$^{-1}$ \\
Efficiency ${\rm{\Phi }}/M_g$              & $5\cdot10^{-9}$yr$^{-1}$      & $10^{-9} - 10^{-8} $yr$^{-1}$ \\
\\
\hline
\\
                                    & \bf{\large {Center Clouds}}    & \bf{\large{Disk Clouds}} \\
\bf{Cloud Characteristics} &  &  \\
\\[-2ex]   \vspace*{3mm}
Mass spectrum:  dN/dM $\propto$M$^{-\gamma}$    & $\gamma \sim$1.6 \,\,\small{($\geq$10$^{4}$ $M_{\odot}$)}                     & 1.6 - 1.7 \\
Vel. dispersion $[$kms$^{-1}]$      & 15 -- 30                            & $\leq$ 5\\
\vspace*{3mm}
Vel.dispersion-Size:    $\bigtriangleup$V [km/s]                 &    12$\cdot$R${^{0.5}}$      &  2$\cdot$R${^{0.5}}$\\
\vspace*{3mm}
Mean Gas Density $[$cm$^{-3}]$       &  $\sim$10$^{{4.0}}$ 	     & $\sim 10^{2.5}$  \\
Temperatures $[$K$]$: Gas                 & {bulk: 50 -- 70}    & $\sim 15$ \\
\qquad\qquad\qquad                  & {wide range of coex. hot phases} & \\
\hspace*{24mm} : Dust \vspace*{3mm}                       & 21 $ {\small\pm2}$       & \\
\vspace*{3mm}
Magnetic Field $[$mG$]$             &  {order mG} & $\leq$ 0.1 \\
Isotopic Abundances                 &                                     & \\
\qquad $[^{12}$C$]/[^{13}$C$]$      & $\sim 25$                           & $\sim 75$\\
\qquad $[^{16}$O$]/[^{18}$O$]$      & $\sim 200$                          & $\sim 400$\\
\qquad $[^{14}$N$]/[^{15}$N$]$      & $\sim 1000$                         & $\sim 400$\\
\\
\hline
\end{tabular}
\caption{Global properties of the central gas layer and cloud characteristics.}
\end{table}
\section{The Galactic Center Magnetosphere}
\label{sec:3}
In the Galactic disk, energy densities associated with the interstellar magnetic field compare with those derived for e.g. the kinetic energies of the
ISM (generally, clouds are observed in approximate energy equipartition). For the enhanced energy densities of the gas in the GC molecular layer,
 equipartition fields of order
 mG are estimated. Observationally though the data base is inconclusive. While locally the dominant impact of magnetic fields on the
 physical processes in the central gas layer is supported by a large number of observations, we have no good handle on the absolute field strength. This
  stems from the general difficulties in observing magnetic fields: while
 we have good tools to derive the
field {\it{morphology}}, the only {\underline{direct}} access to the field {\it{strength}} is via difficult-to-perform line polarization measurements.
\subsection{Magnetic Fields in the Diffuse GC ISM}
\label{sec:1}
The most striking manifestation of the magnetic field in the diffuse ISM (or more precise: in the intercloud medium) are the Non-thermal Filaments (NTFs),
evident in radio continuum images of the Galactic Center \cite{yus84}. These structures - unique to the Galactic Center - are observed isolated or in
bundles of parallel strands,
they run (within 20-30 deg) orthogonal to the Galactic plane for a few tens of parsecs, while their transverse dimension is only a fraction of a parsec.
Their radio spectra with strong linear polarization suggest a non-thermal (synchrotron) nature of the underlying emission process.
Most, if not all, of the NTFs are interacting with
molecular clouds, but reveal little, if any, distortions. The best example is where the {"+25km/s cloud"}, associated with G0.18-0.04, interacts
with the Radio Arc \cite{ser91}. From the filament's rigidity, a magnetic field strength (in the NTF) of a few mG have been estimated by
equating magnetic pressure with the ram and/or intracloud turbulent pressure of
the associated molecular cloud \cite{ser91} \cite{ser94}.   \\
A plethora of theoretical models has been suggested to describe the nature of these filaments, and their relationship with the overall GC field structure
(e.g., \cite{mor96} \cite{mor98}). Common to most of these models is that the NTFs, independent of their very origin, are manifests of
a large-scale poloidal field component. Whether the NTFs are local illuminations of a volume-pervasive (static) strong central dipole field, or whether they
represent local (maybe, dynamic) field enhancements of an otherwise much weaker field, is much of a controversy.
\subsection{Magnetic Fields in the GC Molecular Clouds}
\label{sec:1}
The great advancement of submm/FIR technologies (with large arrays available now for extended mapping) has made possible to perform polarimetry of the
thermal emission from magnetically aligned dust grains in the dense Galactic Center clouds, so providing the field orientation B$_\bot$ (vertical to the line-of-sight).
Fig.4 is a composite of results from polarisation measurements performed over the last couple of years (from \cite{chu03}).
\begin{figure}[hbt]
\centering
\setlength{\unitlength}{1mm}
\begin{picture}(200,75)
\includegraphics{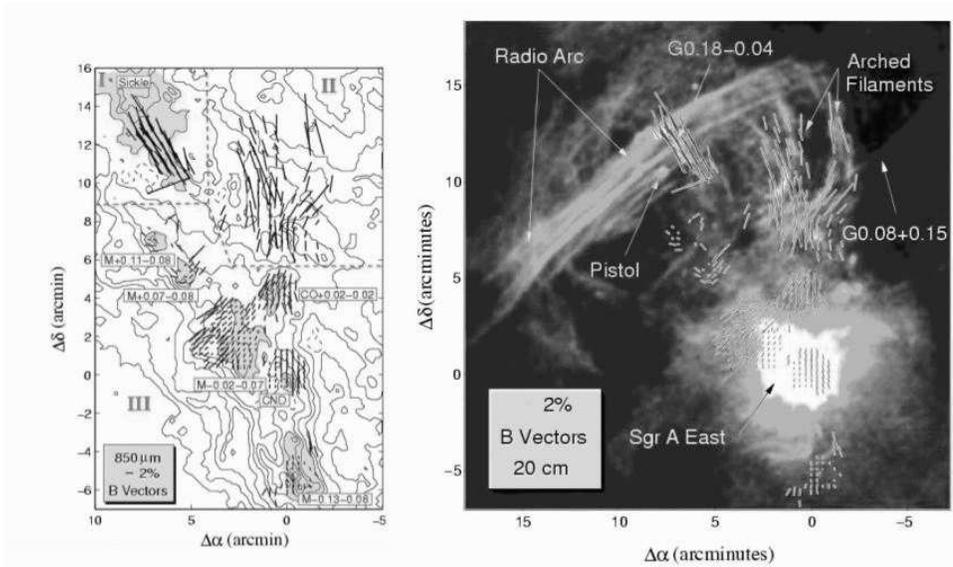}
\end{picture}
\caption{Magnetic field directions as inferred from polarization measurements of magnetically aligned dust grains (from \cite{chu03}), superposed on the 850$\mu$m dust
emission \cite{pie00}. Polarimetry data
are from Dotson et al. \cite{dot00} [region I at 60$\mu$m, region II at 100$\mu$m] and from Novak et al. \cite{nov00} [region III at 350$\mu$m]. Offsets refer
 to  Sgr A$^*$.}
\label{fig:1}       
\end{figure}
\noindent
At first sight, there is little evidence for a poloidal component of the magnetic field associated with the dense clouds: instead, the orientation is
predominantly parallel to the galactic plane (following the main axis of the cloud complexes), more suggestive of a toroidal configuration.
The most amazing discrepancy - in view of their obvious physical interaction - is noticed towards G0.18-0.04, where the field in the
cloud is right perpendicular to the field traced by the NTFs (Fig.4, right panel).
\\
The strong degree of polarization and the rather uniform field orientation within a given cloud complex (as sampled with sub-parsec resolution) suggests a scenario,
in which gravity controls magnetic forces, and differential rotation has sheared the field to its azimuthal configuration \cite{nov00}.
Chuss et al. \cite{chu03},
based on higher resolution data from the Hertz polarimeter on the CSO, report that the field orientation is in fact coupled to the column density of material:
in the dense clouds, gravity dominates and the field is stretched along the long axis of the cloud, whereas in the cloud envelope the field is bending back
to its (original ?) poloidal configuration, perpendicular to the plane. These results, which clearly have to be followed-up and should be complemented by line
polarization measurements \cite{gre02}, that take advantage of the
velocity information along the line-of-sight, seem to provide the clue to understand the apparently contradictory findings of co-existing poloidal and
toroidal field configurations in the center of our Galaxy.
\\
Assuming equipartition conditions in these transition layers, with obvious uncertainties, field strength of order mG are deduced \cite{chu03}. However, as long as we
have no  direct determination (means via Zeeman splitting), the actual strength of the magnetic field must remain uncertain. Except for the circumnuclear disk
(the structure orbiting about the central mass at a few parsec
distance only), for which \cite{kil92}\cite{pla95} report B$_\| \sim$2-3 mG,
no relevant detection has been reported [fields derived from measuremens of OH masers \cite{yus96} compare with findings in the Galactic disk, but because they heavily
reflect local excitation conditions, they shall not be used
to constrain the field in the molecular clouds]. An extended OH absorption survey \cite{uch95} has yielded upper limits only of
$\sim$0.3 mG (3$\sigma$, 13 positions) to any coherent l-o-s field component. While the failure to detect any B$_\|$ component
towards, e.g., G0.18-0.04 (where we have a most convincing, though indirect, case of a mG field) could still be explained by its special field geometry (with no field component
projected towards the observer), statistically these negative results seem to question the existence of pervasive strong magnetic field in the GC clouds.
\section{Perspectives - The Next Steps}
\label{sec:4}
Technological progress has provided us with the ability for wide-field mapping with critically high spatial resolution. A unique data base has been collected, that
 greatly improved
 our understanding of the excitation conditions in the central cloud layer,
but still fundamental questions about the exceptional physics remain to be answered (about the thermodynamics of the gas, its heating processes, the
role of the magnetic fields -- to mention a few). Our review should have affirmed that (1)
understanding the GC is a mandatory (intermediate) step in order to adequately model extragalactic nuclei in general, and (2) that much of the insight derives
from "micro"-physics that will only be possible to resolve in our galactic center due to its close proximity.
\\
By the nature of the underlying physics (the main cooling lines of the cold ISM are in the submm/FIR regime), we do expect critical answers from the new submm and FIR facilities, currently under development.
  APEX and ASTE, ALMA on mid terms, will soon explore the
submm atmospheric windows to the technological limits, with the unique airborne (SOFIA) and space FIR missions (Herschel) already on the horizon.
%
%

%
%



\printindex
\end{document}